# Study of characteristics for heavy water photoneutron source in boron neutron capture therapy

**Danial Salehi**[1*], **Dariush Sardari**[1], **M. Salehi Jozani**[2]

1. *Department of Radiological and Nuclear Engineering, Science and Research Branch, Islamic Azad University, Tehran, Iran.*
2. *Department of Electrical Engineering, South Tehran Branch, Islamic Azad University, Tehran, Iran*

**\* Corresponding Author:** d.salehi@srbiau.ac.ir

**Abstract**:

Bremsstrahlung photon beams produced by medical linear accelerators are currently the most commonly used method of radiation therapy for cancerous tumors. Photons with energies greater than 8-10 MeV potentially generate neutrons through photonuclear interactions in the accelerator's treatment head, patient's body, and treatment room ambient. Electrons impinging on a heavy target generate a cascade shower of bremsstrahlung photons which energy spectrum shows an end point equal to the electron beam energy. By varying the target thickness, an optimum thickness exists for which, at the given electron energy, maximum photon flux is achievable. If a source of high-energy photons i.e. bremsstrahlung, is conveniently directed to a suitable $D_2O$ target, a novel approach for production of an acceptable flux of filterable photoneturons for boron neutron capture therapy (BNCT) application is possible. This study consists of two parts: 1. Comparison and assessment of deuterium photonuclear cross section data. 2. Evaluation of heavy water photonuclear source.

## 1. Introduction

BNCT is a binary radiation treatment that is based on nuclear reactions between thermal/epithermal neutrons and stable isotope $^{10}B$ concentrated primarily in cancer cells. Facilities generating high intensity neutrons find applications in wide-range of fields such as boron neutron capture therapy (BNCT) for cancer treatment, nuclear, materials science, condensed matter physics, polymer science, radioisotope productions, industries, biotechnology, etc. [1-3]

As for BNCT high epithermal neutron flux with energy distribution peaking around 10 keV is needed, a neutron reactor or linear accelerator can be used as the neutron source in BNCT. Linear accelerators are capable of producing neutrons through photonuclear reaction. If the energy of the incident photon is high enough it can interact with the nucleus of an atom and excite it to higher energy state, causing the release of a neutron [4]. Due to the lower energy thresholds of photonuclear interaction in beryllium and deuterium, these two materials are prime choices for photoneutron targets of interest in BNCT. Figure 1 shows this application.

In the photonuclear process high-energy electron beam impinges on a target material; continuous spectrum of bremsstrahlung photons is generated. These bremsstrahlung photons subsequently interact with the nucleus of the target material, resulting in the emission of the nucleons. This interaction is known as photonuclear interaction. The absorption of a photon leads to the formation of a compound nucleus which decays by the emission of one or more neutrons. In order for the neutron to be produced, the absorbed photon must have energy greater than the binding energy of the neutron to the nucleus. This threshold depends on the atomic number of the target: for high atomic numbers it is around 8 MeV whilst for even-even nuclides with low atomic numbers the threshold is higher (16 MeV for oxygen, 18 MeV for carbon).This neutron flux could find application in industrial and medical fields such as BNCT and neutron radiography. Therefore linacs with photon energies in the range of 18–25 MeV can produce undesired fast neutrons [5]. During cancer treatment with medical linear accelerator, neutrons might be produced in the accelerator head, typically collimators, target and flattening filter, and even the patient's body.

As the nucleons are bounded with the nucleus by binding energy (5-15 MeV), the photon should have energy above a threshold value to participate in the photonuclear reaction. Neutrons from the photon-induced giant-dipole-resonance (GDR) reaction consist of a large portion of evaporation neutrons which dominate at low neutron energies (< 1-2 MeV), and a small fraction of direct neutrons, which dominate at high energies [6]. The photoneutron energy spectrum is characterized by an evaporation peak in the range 200–700 keV and a relatively weak (10% of the integrated intensity) direct-reaction component in the several MeV energy range.

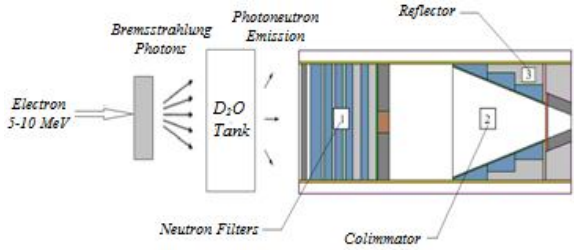

Figure 1: Photoneutron production for BNCT treatment.

## 2. Photoneutron cross section

The photoneutrons are produced isotropically and penetrate the head shielding in all directions. The average energies of the primary neutrons do not vary greatly with peak photon energies, as can be seen from Table 1[4].

Table 1: Average neutron energies produced in linear accelerators.

| Photon Energy [MeV] | Average Neutron Energy [MeV] |
|---|---|
| 15 | 1.8 |
| 20 | 2.1 |
| 25 | 2.2 |
| 30 | 2.4 |

Photoneutron production is dependent on the photonuclear cross section of the material with which the photon interacts. A larger photonuclear cross section means there is a greater probability for the interaction to occur. Photoneutrons are produced in the GDR region, primarily between 3 and 25 MeV, when the incident photon energy is above the production threshold energy. These threshold energies, $Q_n$, are listed in Table 2 for a few nuclides. Threshold energies typically range from 10 to 19 MeV for light nuclei (A < 40) and from 4 to 6 MeV for heavy nuclei.

For most nuclides the binding energy of the highest level neutron, which plays the role of a threshold energy for the photodisintegration reaction, is about 7 MeV [7] and therefore of minor interest in reactor physics. But nature provides some materials that have much lower threshold energies (see Table 2).

Table 2. Nuclides with low photodisintegration threshold energies.

| Nuclide | Threshold(MeV) | Reaction |
|---|---|---|
| $^2$D | 2.225 | $^2$H($\gamma$,n)$^1$H |
| $^6$Li | 3.697 | $^6$Li($\gamma$,n+p)$^4$He |
| $^6$Li | 5.67 | $^6$Li($\gamma$,n)$^5$Li |
| $^7$Li | 7.251 | $^7$Li($\gamma$,n)$^6$Li |
| $^9$Be | 1.667 | $^9$Be($\gamma$,n)$^8$Be |
| $^{13}$C | 4.9 | $^{13}$C($\gamma$,n)$^{12}$C |

Photoneutron emission can occur with all elements as long as the photon energy is greater than the nuclear binding energy of the most loosely bound neutron. The approximate energy of the ejected neutron is given by: [8]

$$E_{pn} \approx \frac{A-1}{A}\left(E_\gamma - Q - \frac{E_\gamma^2}{2m_n c^2 (A-1)} + \frac{E_\gamma}{A}\sqrt{\frac{2(A-1)}{(m_n c^2 A)(E_\gamma - Q)}} \cos\theta\right) \quad (1)$$

where
$A$ is the mass of the target nucleus, $E_\gamma$ is the energy of the photon, $E_{pn}$ is the energy of the neutron, $Q$ is the threshold energy, $m_n$ is the neutron mass, $c$ is the velocity of light, and $\theta$ is the angle between photon and neutron flight direction. The actual useful energy of the photoneutron may be less than that calculated in the above equation due to collisions the neutron may suffer as it travels out of the target.

Photoneutron yields depend upon the target material, the strength and spectrum of the photon source and the geometry. Once the photon source spectrum is determined, the photoneutron yield will be determined by the cross section for the photodisintegration process which is dependent on the photon energy. The photoneutron yield from a target of atomic number Z is given as:

$$Y_n = \int \sigma(Z,G)W(\omega,E)d\omega \quad (2)$$

Integrated over all photon energies above the reaction threshold, where ω is the energy of the photon quanta, σ is the photonuclear reaction cross section; E the energy of the electron and W is bremsstrahlung photons spectrum.

## 2.1. Photoneutron source strength

The neutron source strength is an important quantity for radiation protection. The photoneutron source strength $S_v$ at a given location may be calculated from the energy distribution of the photon flux density by: [7]

$$S_v = \int_{E_t}^{E_{max}} \mu_{\gamma,n}(E)\varphi_y(E)dE \quad (3)$$

where $E_{max}$ is the maximum photon energy, $E_t$ is the threshold for photoneutron production and $\mu_{\gamma,n}$ is the (γ,n) interaction coefficient

## 2.2. Deuterium photoneutron cross section

Deuterium is more sensitive to the energy spread than beryllium. But, at higher energies this energy spread is generally negligible and is only significant when a mono energetic group of neutrons is desired. In heavy water or beryllium-moderator reactors, the photonuclear source may be very appreciable, and the neutron field deep within a hydrogenous shield is often determined by photoneutron produced in the deuterium (which constitutes about 0.015 atom percent of the hydrogen). [9]

Deutron's low photoneutron threshold (2.22 Mev) makes it an important potential neutron source in applications that include heavy water. Because of its light mass, the nuclear theories described in this work are unsuitable for modeling deuteron photodisintegration. Photonuclear cross sections for deuterium from the international nuclear data library EXFOR have been employed as shown in Figure 2.These data were partly compiled into the internationally available computerized library EXFOR (Library of experimental nuclear cross section data in the Exchange FORmat). Table 3 lists the parameters of GDR observed in photo nuclear reaction that has been prepared by IAEA photonuclear data Library. [10]

EXFOR: EXFOR 8-digit entry&subentry number. In Table3, Nucl: Target nucleus (symbol) A: Target nucleus (mass number). Reac: Reaction. $E_{max}$: Energy corresponding to cross section peak. $\sigma_{max}$ : Peak cross section in mb. FWHM: Full width at half maximum in MeV. $E_{int}$: Upper limit of integration in MeV. $\sigma_{int}$: Integrated cross section in MeV*mb $\sigma^1_{Int}$:First moment of the integrated cross section in mb.

Table3. Parameters of GDR observed in photo nuclear reaction cross section for deuterium.

| Nucl | A | Reac | $E_{max}$ Mev | $\sigma_{max}$ mb | FWHM Mev | $E_{int}$ Mev | $\sigma_{int}$ Mev | $\sigma^1_{int}$ mb | Ref. | Author |
|---|---|---|---|---|---|---|---|---|---|---|
| 1-H | 2 | γ,n | 4.48 | 2.5 | 18.1 | 30.00 | 29 | 3.6 | Ann.Phys. 27.79 (1964) | E.G. FULLER |
| | | | | | | 140.0 | 40 | 3.7 | | |

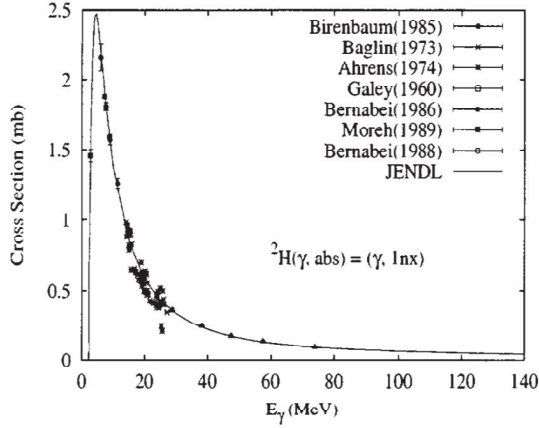

Figure 2. The photonuclear reaction cross sections from the international nuclear data library EXFOR.

Assuming that neutrons and protons are arranged as fermions with spin of 1/2 in discrete potential energy levels, the ground state of the deuterium is known to be a bound triplet $^3S$ state from which a transition into the continuum can take place into the $^0S$ state. This transition involves only a spin flip with change in the spin state of one and is therefore called magnetic dipole (Ml) transition. Higher photon energies enable the transitions from the $^3S$-ground state to the $^3P$-continuum state, which is called an electric dipole (El) transition resulting from a change of the orbital angular momentum of one.

The cross sections for the magnetic and electric dipole transition add up to the total photo disintegration cross section:

$$\sigma_{pn}(E_\gamma) = \sigma_{pn}^{M1}(E_\gamma) + \sigma_{pn}^{E1}(E_\gamma) \qquad (4)$$

The results of computations for photonuclear cross section of deuterium appear in Table 4 and Figure 3 which compare the analytically derived $\sigma_{pn}$ and of $\sigma_{pn}^{M1}/\sigma_{pn}^{E1}$ with some measured values.

Table 4. Computed photoneutron cross section for deuterium.

| E(Mev) | $\sigma_{\gamma n}$(mb) | E(Mev) | $\sigma_{\gamma n}$(mb) | E(Mev) | $\sigma_{\gamma n}$(mb) |
|---|---|---|---|---|---|
| 2.22 | 0.221 | 3.72 | 1.734 | 8.5 | 1.981 |
| 2.32 | 0.322 | 3.82 | 1.834 | 9.5 | 1.852 |
| 2.42 | 0.423 | 3.92 | 1.935 | 10.5 | 1.723 |
| 2.52 | 0.524 | 4.22 | 2.238 | 11.5 | 1.594 |
| 2.62 | 0.625 | 4.32 | 2.338 | 12.5 | 1.466 |
| 2.72 | 0.725 | 4.42 | 2.439 | 13.5 | 1.337 |
| 2.82 | 0.826 | 4.48 | 2.500 | 14.5 | 1.208 |
| 2.92 | 0.927 | 4.5 | 2.497 | 15.5 | 1.079 |
| 3.22 | 1.229 | 5 | 2.432 | 16.5 | 0.950 |
| 3.32 | 1.330 | 5.5 | 2.368 | 17.5 | 0.821 |
| 3.42 | 1.431 | 6 | 2.303 | 18.5 | 0.692 |
| 3.52 | 1.532 | 6.5 | 2.239 | 19.5 | 0.563 |
| 3.62 | 1.633 | 7.5 | 2.110 | 20.3 | 0.434 |

Compared with the existing data, it shows that our calculation is correct. The energy range for this calculation was between 2.22-20.5MeV and the outcome of computation was compared with the literature data set. Among these data for photon energy and photonuclear cross section a minimum in E=2.22MeV and a maximum in E=4.48MeV can be observed.

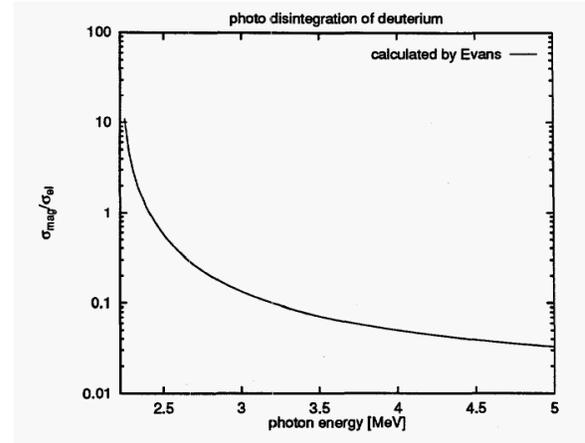

Figure 3. The ratio of the magnetically induced photoneutron cross section to the electrically induced photoneutron cross section $\sigma_{pn}^{M1}/\sigma_{pn}^{E1}$ for deuterium.

## 2.3. Heavy water macroscopic cross section

In this part we examine briefly the cross sections of the heavy water. These data are taken from the MCNP cross section library. As shown in Figure 4, heavy water has a high elastic scattering cross section and a very low absorption cross section at low energies. The elastic scattering cross section of deuterium decreases at energies higher than 2 MeV.

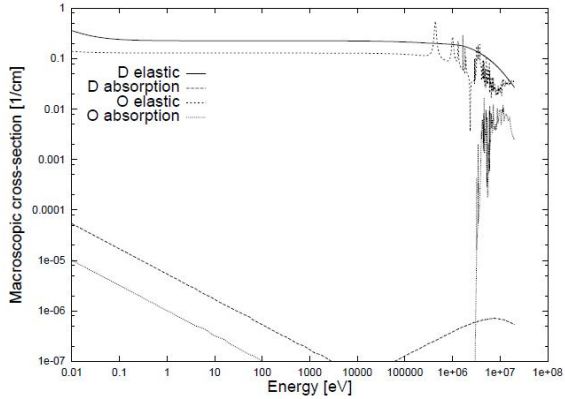

Figure 4. The macroscopic cross sections of heavy water.

## 3. Monte Carlo calculation for photo-neutron and gamma production in $D_2O$

In heavy water and beryllium, respectively, the mean free paths for the Compton effect for photons energetic enough to initiate the photonuclear reaction are approximately 25 cm and 10 cm. So, although a larger target will yield a greater photoneutron production, there exists a point at which the neutron production will peak due to the reduction of the photon flux through other competing interactions.

The following computation shows Monte Carlo code runs using ACCEPT, output from a program written to calculate the photoneutron production within a particular geometry volume.

The ACCEPT code, developed at Sandia National Laboratories, offers a Monte Carlo solution of linear time-independent coupled electron-photon radiation transport problems. Initially, calculations to estimate the photoneutron device's performance were completed using the ACCEPT three-dimensional electron-photon coupled transport code to obtain the bremsstrahlung flux throughout the device.

The ACCEPT example input files given are shown with photoneutron outputs for incident electron energies at 6 MeV and 4 MeV in table 5 and 6 [9]. For the 4 MeV case file and the 6 MeV clinical device case the example source outputs are shown over a heavy water region as well.

In order to produce the neutron beam, the bremsstrahlung from the tungsten targets was directed into sealed cylindrical Lucite tanks of $D_2O$. Although several thicknesses of $D_2O$ cylinders were available, the tank with radius 16 mm and thickness 100 mm surrounded by a thick lead moderator, was found to offer the best neutron production of the available heavy water cylinders, which contained 1920g enclosed in the Lucite tank with a wall thickness of 3.1mm the forward-peaked bremsstrahlung radiation emanating from the target is additionally collimated by a cylindrical tungsten shield. This tungsten collimation was significant and was included in the ACCEPT model of the target apparatus. Using this estimated current, the unfiltered neutron production rate is determined to be on the order of $10^9$ neutrons/second per milliampere of electron beam current. An electron beam in the 4 to 10 MeV range and assuming a 1 mA beam current impinges on the targets yielding forward-peaked bremsstrahlung which is then absorbed in the $D_2O$ photoneutron target volume.

The photoneutrons generated in this volume produce a strong neutron source which then passes through a filtering/moderating region where it is tailored to desired beam specifications. The neutron filter/ moderator should have a high scattering cross section in the high energy range (fast neutrons) and a low cross section for the epithermal neutrons. [11]

Table 5. The computed values of photoneutron production in $D_2O$ for E = 4 and 6 MeV case (N production per mA in volume).

| Incident electron Energy (MeV) | Linac Type | |
|---|---|---|
| | 4(MeV) | 6(MeV) |
| 3.5 | 8.46E+03 | 1.54E+10 |
| 3 | 8.77E+05 | 8.60E+09 |
| 2.5 | 2.76E+06 | 1.08E+10 |
| 2.2 | 1.87E+06 | 4.17E+09 |
| 2 | 0.00E+00 | 0.00E+00 |
| 1 | 0.00E+00 | 0.00E+00 |
| 0.01 | 0.00E+00 | 0.00E+00 |

Table 6. The computed values of photon flux distribution for given target volume in $D_2O$ at E = 4 and 6 MeV.

| Incident electron Energy (MeV) | Linac Type | |
|---|---|---|
| | 4(MeV) | 6(MeV) |
| 3.5 | 2.72E+05 | 1.16E+11 |
| 3 | 3.26E+07 | 3.20E+11 |
| 2.5 | 1.17E+08 | 4.58E+11 |
| 2.2 | 2.80E+08 | 6.24E+01 |
| 2 | 4.74E+08 | 7.74E+11 |
| 1 | 2.82E+09 | 1.44E+12 |
| 0.01 | 1.23E+11 | 6.16E+12 |

## 4. Conclusion

Electron linear accelerators used for medical radiation therapy, operated in bremsstrahlung mode above 10 MV, can produce neutrons through the photonuclear GDR reaction. With 5 MeV electrons, only two materials could be considered, beryllium and deuterium, or some of their compounds, which show a ($\gamma, n$) energy threshold of 1.66 MeV and 2.2 MeV, respectively. The ($\gamma, n$) cross sections for these two materials are of the order of mbarn in the 0-5 MeV energy range. It seems to be a good choice due to the compromise of the low ($\gamma, n$) beryllium threshold and the higher ($\gamma, n$) deuterium cross section.

Capture gamma photons arising from neutron absorption have particularly high energy ($\sim 7\ MeV$) and this may cause a significant production of energetic photoneutrons. The average energy of the generated neutrons from a heavy water moderator is 25 MeV. A photonuclear-based neutron source, with using electron linear accelerators and reasonable yield plus suitable energy spectrum might be applied for BNCT [12].

The proposed photoneutron device could offer a promising alternative approach to the production of epithermal neutrons for BNCT. This technique for the production of epithermal neutrons is not well-developed for clinical applications. Thus mass of data and detailed design of the equipment are still needed. On the other hand, control of photon contamination to acceptable levels at the irradiation point would be crucial to the success of the overall concept. [13]